\documentclass[aps,rmp,preprintnumbers,floatfix]{revtex4} 
\usepackage[dvips]{graphicx}
\usepackage{amssymb}
\usepackage{amsmath}
\newcommand{\be}{\begin{equation*}}
\newcommand{\ee}{\end{equation*}}
\newcommand{\bea}{\begin{eqnarray*}}
\newcommand{\eea}{\end{eqnarray*}}
\newcommand{\bitem}{\begin{itemize}}
\newcommand{\eitem}{\end{itemize}}

\newcommand{\benum}{\begin{enumerate}}
\newcommand{\eenum}{\end{enumerate}}
\newcommand{\bc}{\begin{center}}
\newcommand{\ec}{\end{center}}
\begin{document}
\title{COSMOLOGICAL ACCELERATION: DARK ENERGY OR MODIFIED GRAVITY?}
\author{Sidney Bludman}
\email{bludman@mail.desy.de} \homepage{http://www.desy.de/~bludman}
\affiliation{DESY, 22607 Hamburg, Germany}
\date{\today}
\begin{abstract} The recently observed accelerating 
cosmological expansion or "dark energy" may be either a negative
pressure constituent in General Relativity (Dark Energy) or modified
gravity (Dark Gravity) without any constituent Dark Energy. Low- or
high-curvature modifications of Einstein gravity are distinguished by
the spacetime (Ricci) curvature of their vacua.  If constituent Dark
Energy does not exist, so that our universe is now dominated by
pressure-free matter, Einstein gravity must be modified at low-
curvature, becoming asymptotically de Sitter.

The dynamics of either kind of "dark energy" cannot be derived from
the homogeneous expansion history alone, but requires also observing the
growth of inhomogeneities. Present and projected observations are all
consistent with a small fine-tuned cosmological constant, but also
allow nearly
static Dark Energy or gravity modified at cosmological scales.
The growth of cosmological fluctuations will potentially distinguish
between static and "dynamic "dark energy". But,
cosmologically distinguishing dynamic Dark Energy from Dark Gravity will
require a weak lensing shear survey more ambitious than any now
projected.  Dvali-Gabadadze-Porrati low-curvature
modifications of Einstein gravity may also be detected in refined
observations in the solar system or in isolated galaxy clusters.

We review local and cosmological tests of General Relativity and
modified gravity. Dark Energy is epicyclic in character, requires fine-tuning to
explain why its energy density is just now comparable to ordinary
matter density, and cannot be detected in the
laboratory or solar system. This, along with 
braneworld theories, now motivate searching for Dark Gravity on solar system,
galaxy cluster and cosmological scales.

\end{abstract}

\maketitle \tableofcontents
\section{INTRODUCTION: COSMOLOGICAL SYMMETRY VS. DYNAMICS}

The greatest mystery in cosmology is the ``dark energy'' source of the
late (redshift $z \lesssim 1/2$) cosmological acceleration.  This
``dark energy'' may be static or dynamic and either an additional
negative-pressure matter constituent within General Relativity (Dark
Energy), or a modification of General Relativity (Dark Gravity
\cite{Gu}). This review rests on the observed global homogeneity and
isotropy of the universe (Robertson-Walker cosmology RW), and will
emphasize the difference between Robertson-Walker {\em kinematics} and
{\em dynamics}. We will recall when RW symmetry could determine the
cosmodynamics: To explain the observed present cosmological
acceleration without constituent Dark Energy, Einstein dynamics must
be modified at low spacetime (Ricci) curvature.

Because the homogeneous expansion history $H(z)$ of the global
universe measures only kinematic variables, it cannot fix the
underlying dynamics: cosmographic measurements of the late
accelerating universe, are consistent with either a {\em static}
cosmological constant or a {\em dynamic} ``dark energy'', which itself
may be constituent Dark Energy or modified gravity (Dark Gravity).
The Concordance Model $\Lambda$CDM, with a small {\em static} Dark
Energy or cosmological constant $\Lambda$, gives a good fit to all
present observations, but also allows a moderately dynamic ``dark
energy'' \cite{Spergel}. We will review how static and dynamic ``dark
energy'' and how dynamic Dark Energy and Dark Gravity could be
distinguished theoretically and by future cosmological, local or
intermediate scale observations.

Whether Dark Energy or Dark Gravity, ``dark energy'' has two distinct
dynamical effects: it alters the homogeneous expansion history $H(z)$
and it suppresses the growth of density fluctuations $\delta$ at
large cosmological scale $a(t)$. Because the growth function
$g(z)\equiv\delta/a$ depends on both these effects, large angular scale CMB
temperature anisotropies, the late-time growth of large scale
structure, and refined weak lensing observations potentially
distinguish static from dynamic "dark energy" and Dark
Energy from Dark Gravity (Section II).

In any metric theory of gravitation, the material stress-energy
sources $T_{\mu\nu}$ determines the spacetime (Ricci) curvature tensor
$R_{\mu\nu}$ or Einstein curvature tensor $G_{\mu\nu}\equiv
R_{\mu\nu}-g_{\mu\nu} R/2$.  Before considering dynamical alternatives, we
recall how the asymptotic spacetime curvature, at vanishing matter
density, can
constrain any RW dynamics, without assuming Einstein gravity:
Birkhoff's Theorem is a {\em geometric} theorem, holding in {\em any}
spherically symmetric metric geometry whose vacuum has vanishing
spacetime (Ricci) scalar curvature \cite{Callan,Peebles}.  Applied to
any spatially homogeneous RW universe, Birkhoff's Theorem asserts that
local Newtonian gravity fixes the global dynamics of any
matter-dominated universe whose vacuum is Ricci-flat
\cite{McCrae,Milne,Callan,Weinberg}.  The Ricci curvature of the
vacuum, or cosmological constant,
distinguishes high- from low-curvature alternatives to Einstein's original
gravity theory (Section III.A). High-curvature modifications (such as
\cite{Arkani-Hamed,Randall,Binutray}) require sub-millimeter
corrections to Newton's inverse-square gravity; low-curvature
modifications (such as $\Lambda$CDM, Dvali-Gabadadze-Porrati (DGP)
\cite{DGP,Deffayet2001} theories) can preserve Newtonian gravity
locally, but must be asymptotically dominated by a cosmological
constant.  Without Dark Energy, our
accelerating universe is matter-dominated, and Einstein gravity needs
low-curvature modification. 

Section III also emphasizes that 
contrived
(epicyclic) dynamic Dark Energy can explain the
present acceleration, but still cannot explain the Cosmic Coincidence
(``Why so small now?''), without fine tuning or anthropic reasoning.
This will lead us, in Section V, to consider the Dark Gravity
dynamical alternatives to Dark Energy.

Section IV reviews how Dark Energy or Dark Gravity dynamics determines
the adiabatic and the effective sound speeds, which govern the growth
of fluctuations.  To illustrate how different dynamics and effective
sound speeds can underly the same equation of state, we compare
canonical (quintessence) and non-canonical (k-essence) scalar field
descriptions of Chaplygin gas Dark Energy .

For a truly static cosmological constant, we revert to Einstein's
original definition as an intrinsic {\em geometric} classical
parameter. By disconnecting the cosmological constant from energy-momentum sources,
we side-step the mysteries of why quantum vacuum fluctuations
apparently do not gravitate and why the present matter density is
rougly equal to the present ``dark energy'' density.  We review
present local and cosmological constraints on General Relativity,
before proceeding to prospective cosmological, solar system and
isolated galaxy cluster tests for modified gravity. We emphasize the
difficulties coming tests of relativistic cosmology face: The next
decade may distinguish static from dynamic "dark energy", but will
still not distinguish constituent Dark Energy from Dark Gravity
\cite{Ishak}. Besides cosmological tests, low-curvature modifications
of Einstein gravity may yet be tested in the solar system (anomalous
orbital precession, increasing Astronomical unit) or in any isolated
rich cluster of galaxies \cite{LueStarkman,Iorio,IorioSecular}
(Section V).

In conclusion, cosmological scale modifications of classical Einstein
gravity are less contrived than fine-tuned Dark Energy and
arise naturally in braneworld cosmology.  By making intrinsic
curvature the source of cosmological acceleration, Dark Gravity avoids
an additional epicyclic matter constituent, may unify early
and late inflation, and may be refuted by
laboratory, solar system, or galaxy cluster (Section VI).

\section{EXPANSION HISTORY H(z) IN ROBERTSON-WALKER COSMOLOGIES}

Our universe is apparently homogeneous and isotropic
(Robertson-Walker) in the large.  These Robertson-Walker
cosmologies are four-dimensional conformally-flat generalizations of
General Relativity, in which the spacetime (Ricci) curvature $R$ and
the Einstein scalar curvature $G=3(k/a^2+H^2)$ are determined by the matter density $\rho$, according to
the gravitational field equations. The homogeneous expansion of our
flat Robertson-Walker universe is described by the kinematic
(geometric) observables in Table I, wherein the cosmological scale
$a(t)=1/(1+z)$ and the number of e-folds $N\equiv \ln a$, so that
$dN=-d\ln(1+z)=Hdt=\mathcal{H}d\eta$. Overhead dots denote
derivatives with respect to cosmic time $t$, so that the conformal
Hubble expansion rate $\mathcal{H}\equiv\dot{a}$, the Hubble
expansion rate $H\equiv \dot{a}/a$, and the Hubble time $H^{-1}=d
~d_M/c dz$  
is the derivative of the conformal time $d_M(z)/c$ back
to redshift $z$. Subscripts $0$ denote present values, so that
$H_0=73\pm 3 ~km/sec/Mpc,~H_0^{-1}=13.4\pm 0.6~Gyr,~c H_0
^{-1}=4.11\pm 0.17 ~Gpc$ \cite{Spergel}.

By measuring the evolution of the mean curvature of the background,
cosmography maps the homogeneously expanding universe,
without reference to dynamics or sources of curvature. However, we
will see in Section III.C that the asymptotic Ricci curvature or
vacuum Ricci curvature $R_{\infty}\equiv
4\varkappa^2\rho_{DE}(a=\infty)$ does constrain the Robertson-Walker
cosmodynamics, distinguishing high- and low-curvature modifications of
Einstein gravity. In Einstein gravity, 
\be G=\varkappa^2\rho/3, \varkappa^2\equiv 8\pi G_N \equiv 1/M_P^2
,\ee
in terms of Newton's constant $G_N$ and the reduced Planck mass $M_P$.
In Einstein's original field equations, 
 $H(t)$ is the only degree of freedom,
only the tensor components of the metric $g_{\mu\nu}$ are
propagating, and (absent a cosmological constant) the asymptotic or
empty space scalar curvature $R_{\infty}=0$.  When Einstein gravity is modified, $\dot{H}\equiv d H/d
t$ and $\ddot{H}\equiv d^2 H/d t^2$, or the cosmological acceleration
$q(t)$ and jerk $j(t)$ become
additional degrees of freedom, describable by scalar or vector
gravitational fields.

Conformal flatness means that light propagates in  
Robertson-Walker cosmologies as in Minkowski space. This directly
implies a Hubble expansion in cosmological scale $a(t)$, an expansion
history $H(z)$, different cosmological distances, and other kinematic
quantitities listed in Table I.

\begin{table*}   
\caption{Kinematic observables for any RW geometry, in terms of Hubble expansion rate $H
  \equiv \dot{a}/a$. }
\begin{ruledtabular}
   \begin{tabular}{|l||r|}
Description                           &Definition            \\
\hline \hline
Hubble horizon                        &$1/\mathcal{H}\equiv 1/aH =d\eta/dN$\\
bulk expansion                        &$da^3/a^3=3dN=3Hdt=3\mathcal{H} d\eta$\\
conformal time since big bang &$\eta(z)\equiv
\int_0^t dt'/a(t')=\int_z^{\infty}dz'/H(z')$ \\
proper motion distance back to redshift z &$d_M(z)=c\int_0^z dz'/H(z')=c(\eta_0-\eta(z))$\\
spacetime curvature                   &$R\equiv 6(\dot H +2
H^2)=6 H^2(1+q)$ \\
\hline acceleration &$\ddot{a}/a=H^2+\dot{H},\qquad
q\equiv\ddot{a}/a H^2=1-dH^{-1}/dt=-d(1/\mathcal{H})/d\eta\equiv
1-\epsilon_H$ \\
"slow-roll" parameter                 &$\epsilon_H\equiv dH^{-1}/dt=-d\ln
H/dN$ \\
overall "equation of state            &$w(z)=d\ln {H^2/(1+z)^3}/3 d\ln
{(1+z)}=-1+2\epsilon_H/3$ \\
cosmological jerk                     &$\dddot{a}/a=H^3+3 H
\dot{H}+\ddot{H},\qquad j\equiv\dddot{a}/aH^3=1+3\dot{H}/H^2+\ddot{H}/H^3$\\
   \end{tabular}
 \end{ruledtabular}
\end{table*}
\subsection{Kinematics: Distances to Supernovae, Luminous Red Galaxies, Last
  Scattering Surface}

The CMB shift and the first baryon acoustic peak are standard rulers
measuring proper motion distances \be d_M(z)\equiv\int_0^z
dz'/H(z')\ee back to the last scattering surface at redshift
$z_r=1089$ and to luminous red galaxies typically at redshift
$z_1=0.35$, by observing the CMB shift parameter
$S\equiv\sqrt{\Omega_{m0}} H_0 d_M(z_r)=1.716\pm 0.062$
\cite{Spergel})and the first baryon acoustic peak
$A\equiv  \sqrt{\Omega_{m0}} H_0 [d_M^2 (z_1)/z_1^2
H(z_1)]^{1/3}=0.469$  \cite{Fairbairn,Eisenstein}. Calibrated supernovae Ia are
standard candles at low redshift $z<1.7$, whose observed
flux=absolute luminosity$/4\pi d_L(z)^2$, measures their luminosity
distance $d_L(z)=(1+z) d_M(z)$ \cite{SCP,Riess,SNLS}.

These cosmological distances then map the evolution history $H(z)$,
but only after differentiation with respect to redshift.
Differentiating the Gold Sample \cite{Riess} supernova luminosity
distances $d_L(z)$, Tegmark {\it et al} \cite{Tegmark4} obtained
Figure 1, a plot of $h(z)\equiv H(z)/100$. A second differentiation
of the observed distances gives the overall "equation of state"
$w(z)\equiv\gamma(z)-1\equiv d\ln{H^2 /(1+z)^3}/3d \ln{(1+z)}$,
decreasing from $0$ in the matter-dominated epoch, to $-2/3$ at
present, and apparently tending towards $-1$ in the far future
\cite{Tegmark1,Tegmark3}. Determining the overall effective
``equation of state'' thus requires two numerical differentiations
of the sparse, noisy primary data on cosmological distances.

As will be discussed in Section III.A below, the late accelerating
expansion is conventionally described by an effective mixture of
General Relativity ideal fluids, now consisting of pressure-free
matter (baryons+CDM) and "dark energy" defined by $\rho_{DE}\equiv 3
M_P^2 H^2-\rho_m $ and $\gamma_{DE} \equiv -d \ln{\rho_{DE}}/3 dN $,
so that the overall stiffness and ``equation of state'' \bea
\gamma\equiv -d\ln{H^2}/3 dN=\gamma_m \Omega_m +
\gamma_{DE}\Omega_{DE},\qquad \Omega_{DE}\equiv 1- \Omega_m
\\w=w_m\Omega_m+w_{DE}\Omega_{DE} = w_{DE}(1-\Omega_m).\eea If Dark
Energy exists as a matter constituent in General Relativity, then
$w_{DE}$ is its "equation of state".  Otherwise, $w_{DE}
(1-\Omega_m)=d \ln{H^2/(1+z)^3}/3 d \ln{(1+z)}$ measures the Dark
Gravity modification to the Einstein-Friedmann equation.

Because $\Omega_{m0}\sim 1/3$, the effective ``dark energy equation
of state'' is now $w_{DE0}\sim -1$, so that the ``dark energy'',
whether Dark Energy or Dark Gravity, is now static or quasi-static,
nearly a cosmological constant $\Lambda
\equiv\varkappa^2\rho_{DE}\approx 2 H_0^{-2}$. Indeed, the recent
three-year WMAP data, with narrowed constraints on $\Omega_m
h^2=0.126 \pm 0.009,~\Omega_m=0.234\pm0.035, ~n_s=0.961\pm
0.017,\text{expansion age} ~t_0=13.73^{+0.13}_{-0.17} ~Gyr$, large
scale structure \cite{Tegmark2004} and supernova data, makes
$w_{DE0}=-0.926^{+0.051}_{-0.075}$ \cite{Spergel}, consistent with the
Concordance Model $\Lambda$CDM and severely limiting
how dynamic any ``dark energy'' can now be.

\begin{figure}  
\includegraphics[width=1.0\textwidth]{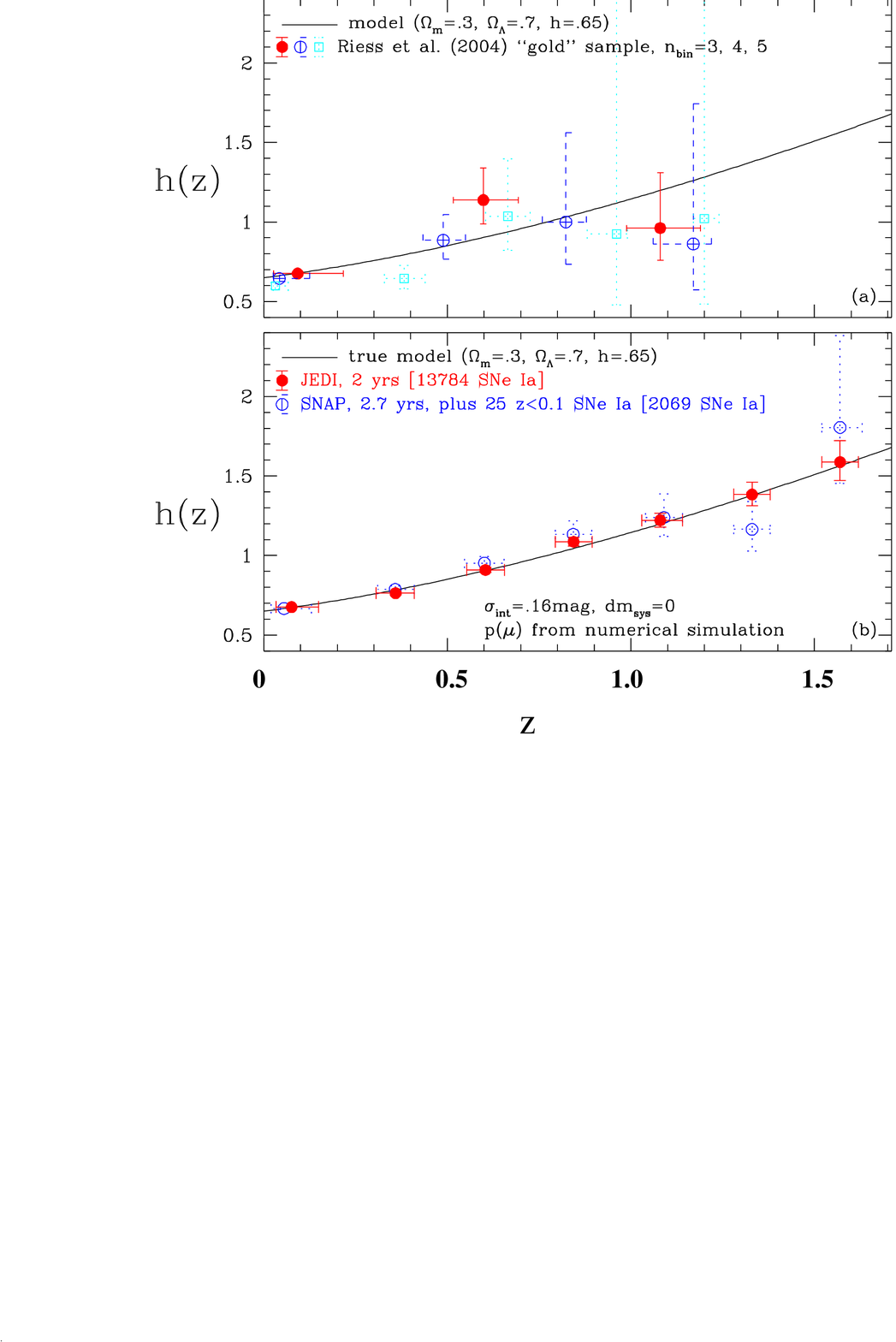}
\caption{The cosmic expansion history (dimensionless Hubble
parameter h(z)=H(z)/100) from the Riess {\em et al.} (2004)
 Gold sample (top panel) and from simulated future data
 (bottom panel) for the NASA/JDEM mission concepts JEDI
 (solid points) and SNAP (dotted points) \cite{Tegmark4}.
 The existing Gold sample data and the simulated future JEDI
 are both consistent with $\Lambda CDM$ (solid curve)
 (from \cite{Tegmark4}).}
\end{figure}

\subsection{Expansion History Does Not Determine Cosmodynamics} 

The expansion history $H(z)$ constrains but does not determine the
cosmodynamics.  Most simply, in Einstein's field equations \be
G_{\mu\nu}=\varkappa^2 T_{\mu\nu}/3, \qquad \varkappa^2\equiv 8\pi
G_N\equiv 1/M_P^2, \ee whose time-time component is the
Einstein-Friedmann equation \be k/a^2+H^2=\varkappa^2 \rho/3, \ee the
dynamics and expansion history cannot distinguish between a
cosmological constant (static Dark Gravity) $~-\Lambda g_{\mu\nu}$
added to the left side and a constant vacuum energy density (static
Dark Energy) $\rho_{vac}=\Lambda g_{\mu\nu}/\varkappa^2$ on the right
side. This familiar Dark Energy/Dark Gravity degeneracy persists when
both are dynamic.  Thus, Table II presents five two-parameter fits to
the combined Supernova Legacy Survey \cite{SNLS}, baryon acoustic peak
and CMB observations \cite{Maartens}. The first and best fit is to the
Concordance Model \be (H/H_0)^2=\Omega_{m0}(1+z)^3+\Omega_{\Lambda
  0}+\Omega_{K0},\qquad \Omega_{m0}+\Omega_{\Lambda
  0}+\Omega_{K0}\equiv 1.\ee The next two fits are to spatially flat
Dark Energy models, in which the Einstein-Friedmann equation is \be
(H/H_0)^2=\Omega_{m0}(1+z)^3+\Omega_{X0}(1+z)^{3(1+\overline{w_{DE}})},\qquad
\Omega_{m0}+\Omega_{X0}\equiv 1, \ee and the "equation of state" is
$w_{DE}=const$ or $w_0+w_a z/(1+z)$, with past-average
$\overline{w_{DE}}\equiv (1/N)\int_{0}^{N}w_{DE}(N')\,dN'=w_{DE}=
w_0-w_a z \ln{(1+z)}/(1+z)$ \cite{Linder2005}. The last two fits
\cite{SNLS,Maartens} are to the original spatially curved
Dvali-Gabadadze-Porrati (DGP) Dark Gravity model
\cite{DGP,Deffayet2001,Bentofit} \be
(H(z)/H_0)^2=[1/2\beta+\sqrt{(1/2\beta)^2+\Omega_{m0}
  (1+z)^3}]^2+\Omega_{K0}(1+z)^2,\quad \Omega_{m0}+\Omega_{K0}
+\sqrt{1-\Omega_{K0}}/\beta \equiv 1,\ee and to a generalized flat DGP
model \be (H/H_0)^2=\Omega_{m0} (1+z)^3 + (1-\Omega_{m0})
(H/H_0)^{\alpha},\ee fitted only to SNLS+BAO and assuming the prior
$\Omega_{m0}=0.27^{+0.06}_{-0.04}$ \cite{Fairbairn,Sahni}. The spatial
curvature $\Omega_{K0}$ is set equal to zero in the
\cite{SNLS,Linder2005,Fairbairn} models and emerges as practically
zero in the \cite{Maartens} models. The dynamical Dark Energy and Dark
Gravity expansion histories are equivalent under the substitution
$(1+z)^{3(1+\overline{w_{DE}})} \leftrightarrow (H/H_0)^{\alpha}$,
with an average $\overline{w_{DE}}=-1+\alpha/2=-1.08^{+0.44}_{-0.32}$
\cite{DvaliTurn}. The Gold SN data \cite{Riess} would have yielded
slightly poorer fits than the SNLS data we have used.

\begin{table*} 
\caption{Model Fits to the Observed Expansion History: Two parameter fits to the joint SNLS,
BAO and CMB shift data by the Concordance Model, two
different dynamical Dark Energy, and two different Dark Gravity
cosmological models.}
\begin{ruledtabular}
   \begin{tabular}{|l||l|l|r|}
Model                                              &$\Omega_K$&$\Omega_{m0}$  &other parameters fitted\\
\hline \hline
$Concordance ~Model~(\Lambda$CDM) \cite{Maartens}    &-0.0050   &0.265           &$\Omega_{\Lambda}=0.740$\\
\hline
flat constant $w_{DE}$ \cite{SNLS}                 &0         &$0.271\pm 0.021$&$w_{DE}=-1.023\pm 0.087$\\
flat $w_{DE}(z)=w_0+w_a z/(1+z)$ \cite{Linder2005} &0         &0.260           &$w_0=-0.78,\qquad w_a=0.32$\\
\hline
original Dvali-Gabadadze-Porrati \cite{Maartens}   &-0.0297   &0.260           &$\beta\equiv H_0 r_c=1.39$    \\
generalized flat DGP \cite{Fairbairn}              &0         &$0.27^{+0.06}_{-0.04}$&$\alpha=-0.17^{+0.87}_{-0.63}$\\
   \end{tabular}
 \end{ruledtabular}
\end{table*}

Because the supernovae are observed only at low redshifts and the CMB
first acoustic peak and the luminous red galaxies at recombination
redshifts $z_r=1089$ and $z_1=0.35$, other smooth parameterizations
could fit the past data equally well. In any case, because the
cosmological distances involve two integrations over $w(z)$, they
all smear out information on the "equation of state" \cite{Moar}. This requires
smooth parametrization of the "equation of state" and binning of the
sparse, noisy data \cite{Tegmark3}, and justifies using no more than
two parameters for present and next decade observations
\cite{LindHut,CaldLinder}. (In retrospect, because observations
constrain the directly observable $H^2(z)$ and the "dark energy"
density better than its derivative, the "equation of state", it
might have been better to parameterize the past average
$\overline{w_{DE}}(z)$,
rather than $w_{DE}(z)$ \cite{WangFreese} ).

The observed evolutionary history is already somewhat better fitted
by the static $\Lambda$CDM model \cite{Maartens} than by any
dynamical Dark Energy or Dark Gravity model in Table II, and the
latest WMAP data  $w_{DE0}=-0.926^{+0.051}_{-0.075}$ \cite{Spergel},
even more
severely limits how dynamic any ``dark energy'' can now be.
Nevertheless, the uncertainties still allow some late-time 
evolution of $w(z)$. Our purpose will now be to discriminate among these nearly
static alternatives by observing the fluctuation {\em growth factor}
on Hubble horizon scales. Figure 2, from \cite{Tegmark1}, shows the
ranges of redshift and conformal length scales over which such
spacetime fluctuations are likely to be measured cosmologically, in
the next few years.
\begin{figure}
\includegraphics[width=1.0\textwidth]{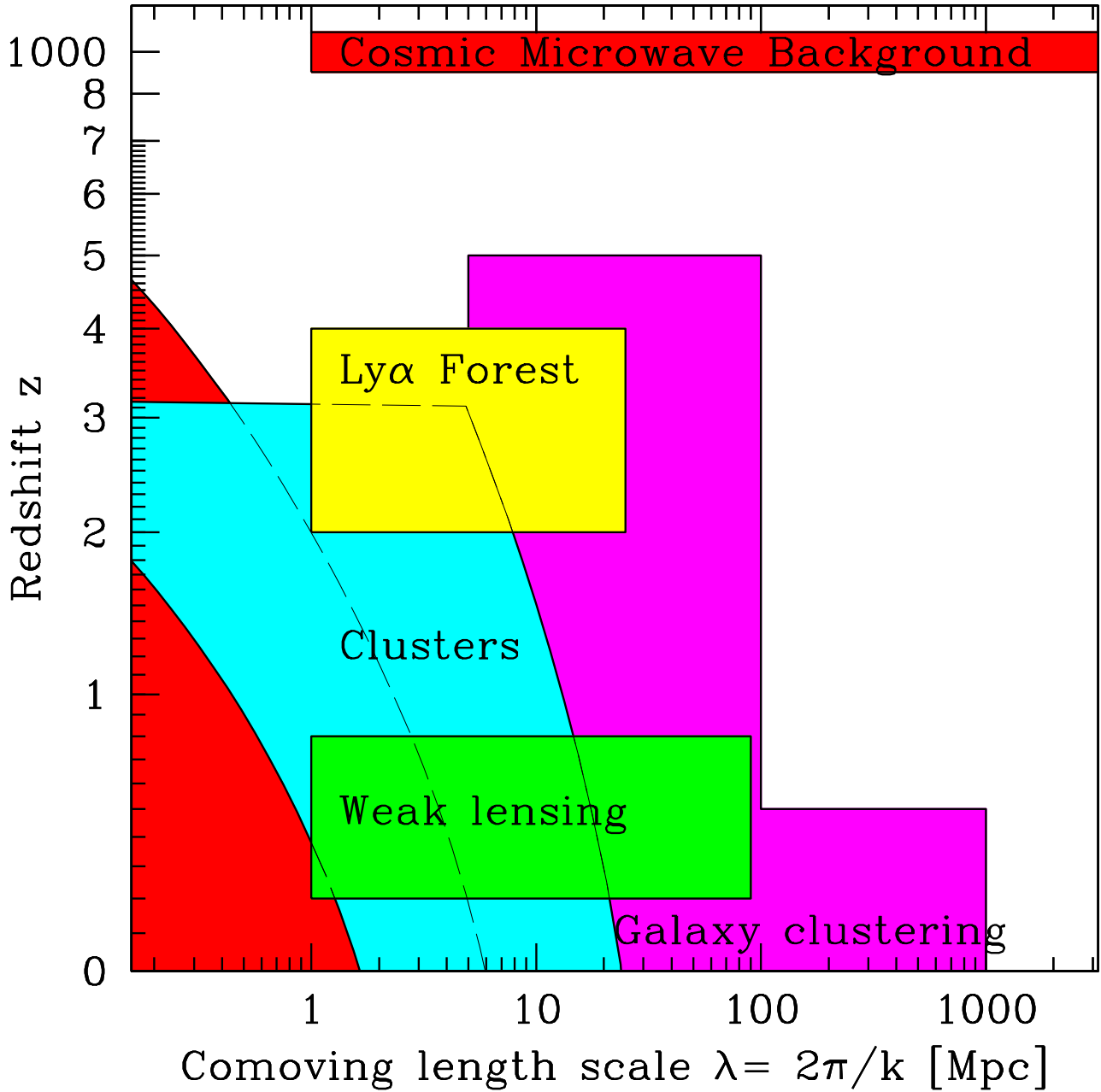} 
\caption{Shaded regions show ranges of scale and redshift over which
various cosmological observations are likely to measure spacetime
fluctuations over the next few years.  The lower left region,
delimited by the dashed line, is the non-linear regime where rms
density fluctuations exceed unity in the $\Lambda$CDM model (from
\cite{Tegmark1}).}
\end{figure}

\section{CLASSIFICATION OF ROBERTSON-WALKER COSMOLOGIES BY THEIR
  VACUUM SYMMETRY}

\subsection{Homogeneous Evolution Conventionally Described by General Relativity Perfect Fluids}

By General Relativity, we understand Einstein's original field
equations without cosmological constant, before
introducing the cosmological constant on the left (geometric) side
of his original field equations, changing the field equations from
Einstein to Einstein-Lemaitre and the asymptotic Ricci curvature from
flat to curved. This introduces into the classical action a very small
energy scale $\Lambda\sim H_0^2\lll M_P^2$.  By avoiding identifying
the cosmological constant with any right-side matter stress-energy
content, this classical approach distinguishes static from dynamic
``dark energy'' and avoids considering the two cosmological constant
problems, why quantum vacuum energies apparently do not gravitate and
why the present matter density $\rho_{m0}\sim M_P^2~\Lambda$.
(Replacing the Einstein Lagrangian $R$ by $R-2\Lambda$ is equivalent to
unimodular gravity \cite{Buch,Unruh} in which the Einstein-Hilbert
action is varied holding $\sqrt{g}=-1$: Instead of appearing in the
Hilbert-Einstein action, $\Lambda$ then enters as an undetermined
Lagrange multiplier.)

Although Robertson-Walker cosmology does not assume General
Relativity, its expansion history may be expressed in terms of an equivalent
perfect fluid defined by $\rho \equiv 3
M_P^2 H^2\equiv\rho_m+\rho_{DE}$ and

\begin{description}
\item [overall equivalent pressure:]$P/c^2 \equiv -M_P^2(3
  H^2+2\dot{H})\equiv P_m +P_{DE}$
\item [overall "equation of state":]$w\equiv P/\rho
  c^2=-(1+2\dot{H}/3 H^2)\equiv -1+2\epsilon_H/3=w_m\Omega_m+w_{DE}\Omega_{DE}$
\item [overall equivalent fluid stiffness:]$\gamma (z)\equiv
  1+w=-2\dot{H}/3 H^2 \equiv 2\epsilon_H/3=\gamma_m\Omega_m+\gamma_{DE}\Omega_{DE}$
\item [overall equivalent enthalpy density:]
  $\rho+P/c^2\equiv -2 M_P^2 \dot{H}=-d\rho/3 H
  dt=\rho_m+P_m-d\rho_{DE}/3 dN.$
\end{description}
So defined, $\rho_{DE}$ is either constituent Dark Energy or a Dark Gravity
addition to
the Einstein-Friedmann equation that is non-linear in $H^2$.
For ordinary non-relataivistic matter $\rho_m\sim a^{-3},
\gamma_m=1,P_m=0=w_m$. Thus, ever since matter dominated over
radiation, the
overall "equation of state" $w=w_{DE}(1-\Omega_{m0}).$

Our universe apparently evolved from a high-curvature de Sitter
(early inflationary) universe $P=-\rho=const$. Assuming the Weak
Energy Condition $w\geq -1$, so that no
phantom matter or cosmic rip intervenes, it will expand
monotonically $\dot{H}\leq 0$, towards a different (late inflationary) low-curvature
de Sitter universe. The deceleration $-q(t)$ decreased from $1$,
when radiation dominated, and is still decreasing, along with the
Hubble expansion rate and Ricci curvature.

Deceleration changed over to acceleration when the Hubble horizon changed
from expanding to shrinking with conformal time, when the
"slow-roll" parameter $\epsilon_H\equiv dH^{-1}/dt=-d\ln H/dN$ fell
below $1$, and $w$ fell below $-1/3$. The acceleration has already
increased to present values $q_0\approx 0.52,~w_0\approx -0.74$, but
because $\epsilon_{H0}\approx 0.4$, this recent inflation is still
far from truly slow-rolling. The jerk $j(t)$ increases monotonically
from a minimum $j_{min}=-1/8$ when $w=-1/2$, towards $j \rightarrow
1$, as the universe asymptotes towards a terminal de Sitter
universe, with low constant curvature $R_{\infty}=12
H_{\infty}^2\sim 4 H_0^2.$ Such a de Sitter attractor explains the
growth of "dark energy", but not why it {\em now} approximates the
energy density of ordinary matter.

During three long epochs listed in Table III, the universe is
dominated by a single barotropic phase with a constant equation of
state $w$ and diminishing Ricci scalar curvature $R(t)$, in which
the scale $a(t)\sim t^{2/3(1+w)},~H=2/3(1+w)t$ and the acceleration
and jerk are fixed at $q=-(1+3w)/2,~j=1+9w(1+w)/2$. When these
perfect fluids are mixed or when cosmological scalar fields appear,
the "equation of state" $w(z)$ and the jerk
$j(z)=1+9w(1+w)/2-3dw/2dN$ change, the composite fluid is imperfect
and supports entropic perturbations.

\begin{table*}   
\caption{Kinematics and Ricci curvature for barotropic phases with power-law growth $a\sim
t^{2/3(1+w)}, ~a\sim\exp(Ht)$.}
\begin{ruledtabular}
   \begin{tabular}{|l||c|c|c|c|c||r|}
w      &a(t)            &H(t) &q(t) &j(t)&$R(t)$      &Model Universe\\
\hline \hline
1/3    &$t^{1/2}$ &1/2t &-1   &3   &0                 &radiation \\
0      &$t^{2/3}$ &2/3t &-1/2 &1   &3$H^2$            &non-relativistic matter (Einstein-de Sitter)   \\
-1/3   &$t$       &1/t  &0    &0   &6$H^2$            &curvature
energy (constant conformal expansion$\mathcal{H}$, Milne\\
-1     &$\exp{Ht}$&const &1   &1   &12$H^2$           &empty de Sitter\\
   \end{tabular}
 \end{ruledtabular}
\end{table*}

\subsection{Dark Energy Requires Fine Tuning to Explain the Cosmic Coincidence}
Dark Energy is reviewed in \cite{Pad}. If it exists, Dark Energy is usually attributed to an additional
ultra-light scalar field $\phi$, so that $p=p(\rho,\phi)$
is adiabatic only when the scalar field is frozen or tracks the
background. Defining $X\equiv \partial_\mu{\phi}\partial^\mu{\phi}$, the scalar field is canonical ({\em quintessence}) when
its kinetic energy is  
$X/2$, non-canonical
({\em k-essence}) when the kinetic energy is non-linear in $X$.
Canonical quintessence is driven by a slow-rolling potential and can
track the background matter, making $dw/dz>0$. k-essence is driven
by a non-canonical kinetic energy and can be arranged to switch from
tracking during radiation dominance over towards a dominating
cosmological constant, making $dw/dz<0$ recently. Of course, this
different dynamics gives quintessence and k-essence different
clustering properties.

The present ``dark energy'' density $\rho_{DE0}\sim
2 H_0^2 M_P^2 \lll M_P ^4$.  While ultra-light dynamic Dark Energy can
evolve down to this very small value, it does not explain the Cosmic
Coincidence, "Why so small now?", why $\rho_{DE0}\sim 2\rho_{m0}$, without
the extreme fine-tuning it was invoked to avoid. Canonical and
non-canonical Dark Energy ultimately require fine-tuning:
quintessence, in order to explain the Cosmic Coincidence; k-essence,
in order to initiate the transition towards a cosmological constant
after radiation dominance ends.
\subsection{Without Dark Energy, Our Universe Must Be Asymptotically
  de Sitter}

{\em Birkhoff's Theorem} \cite{Callan,Peebles}: In any locally
isotropic (spherically symmetric) system whose vacuum is Ricci-flat
($R_{\mu\nu}$=0), the vacuum metric must be Schwarzshild: \be
g_{tt}=g_{rr}^{-1}=1-2G_NM(r)/r,\ee where $M(r)$ is the mass interior
to $r$.  For any small spherical shell in empty space, the Newtonian
potential must vanish inside, and decrease as $1/r$ outside.
Birkhoff's Theorem is a {\em geometric} theorem, which generalizes
Newton's iron sphere theorem \cite{Callan,Peebles} from Newtonian
gravity to Einstein gravity or any high-curvature modification of
Einstein gravity.

{\em Application to any Robertson-Walker
  Cosmology} \cite{Callan,Weinberg}: Applied to a homogeneous universe
with matter density $\rho(a),~M(r)=4\pi\rho(a) r^3 /3$,
Birkhoff's Theorem has remarkable {\em dynamical} consequences. In a
homogeneous expanding universe, a small comoving shell lying at
$r(t)=\lambda_{*} a(t)$, encloses a mass $M(r)=\lambda_{*}^3 4\pi
\rho(a) a^3/3$, and has constant Newtonian energy \be \dot{r}^2
/2-G_NM(r)/r=\lambda_{*}^2[\dot{a}^2/2 +\varkappa^2\rho(a) a^2/6] .\ee
Using Birkhoff's Theorem, Milne and McCrae \cite{McCrae,Milne}
derived the global Friedmann equation \be \dot{a}^2-\varkappa^2\rho
a^2/3=const,\qquad k/a^2+H^2= \varkappa^2\rho/3\ee for any pressure-free universe, without assuming Einstein's field equations. If
Dark Energy does not exist and the vacuum is Ricci flat, Birkhofff's
Theorem would make our presently pressure-free universe
{\em decelerate} according to this Friedmann expansion equation.

{\em Classification of Dark Gravity Theories:} If there
is no Dark Energy, the Ricci curvature of the vacuum distinguishes
high-curvature from low-curvature modifications of General
Relativity: Robertson-Walker universes whose vacua remain Ricci-flat
in four dimensions (e.g. Arkani-Hamed {\it et al}
\cite{Arkani-Hamed}, Randall-Sundrum \cite{Randall}, Binutray
\cite{Binutray}) can only modify Einstein gravity in the
ultra-violet. Robertson-Walker universes which maintain Einstein
gravity locally can only modify Einstein gravity cosmologically
(e.g. $\Lambda$CDM, self-accelerated DGP).  If there is no Dark
Energy, our accelerating universe is now dominated by pressure-free
matter, so that the Friedmann-Robertson-Walker equation must be modified
at {\em low} Ricci curvature, by introducing a cosmological constant.

{\em When Einstein gravity is modified:} 
If the universe is asymptotically Ricci-curved, the modified
Friedmann equation \be k/a^2+H^2=\varkappa^2 f(\rho)/3, \ee
maintains Einstein gravity at high density $f(\rho)
\rightarrow\rho$, but crosses over to de Sitter $\rightarrow const
\equiv M_P^2 ~\Lambda $ at scales $r_c$ approaching the Hubble
horizon $H(z)^{-1}$. About any isolated source of mass $M$ and Schwarzchild
radius $r_S\equiv 2 G_NM$, Einstein gravity remains a good
approximation only for distances $r<r_{*}$ up to the
Vainstein scale $r_{\star}\equiv (r_S r_c^2)^{1/3}\sim (H_0
r_S)^{1/3} H_0^{-1}\ll H_0^{-1}$ at which $2 G_N M/r_{*}\equiv
r_S/r_{*}=H^2(a) r_{*}^2= r_{*}^2/r_c^2$. These are the promising
low-curvature modification of Einstein gravity that asymptote to de
Sitter, to be discussed
in Section V.B.
\section{DYNAMICS DETERMINES THE GROWTH OF FLUCTUATIONS} 

Allowing inhomogeneities breaks translational invariance, leading to
Goldstone mode sound waves that lower the CMB angular power spectrum
at large scales (low multipoles $l$) and leads to growth of large scale
structure. In a mixture of cosmological fluids or dynamic scalar
fields, the equation of state is generally not adiabatic: fluctuations
propagate in the conformal Newtonian gauge with an {\em effective
  sound speed} $c_s^2=P_{,X}/\rho_{,X}=w-d w/3(1+w) dN)$, generally
different from the adiabatic sound speed $c_a^2=\partial P/\partial
\rho=\dot{P}/\dot{\rho}$. For canonical scalar field quintessence,
$c_s^2=1$, but for non-canonical k-essence, $c_s$ can vary rapidly,
nearly vanishing near the radiation/matter cross-over, where $w(z)$ is
changing.

Because the
entropic pressure fluctuations are proportional to
$(1+w)(c_s^2-c_a^2)$, they are insensitive to the effective sound
speed in the quasi-static limit $w(z) \sim -1$. This minimizes the
differences between static and dynamic "dark energy" and between any
dynamical Dark Energy and Dark Gravity, making their fluctuation
growth factors hard to
distinguish, in present and in next-generation experiments
(Section V.C).

With the same equation of state and adiabatic sound speed, different
dynamics generally leads to different effective sound speeds. This
equation of state degeneracy is illustrated by the toy Chaplygin
Gas, whose adiabatic equation of state $P=-A/\rho$ and sound speed
$c_a^2=-w(z)$ can be derived from adiabatic fluid dynamics, from a
non-canonical Born-Infeld scalar field, or from a canonical tracking
scalar field with potential
$V(\phi)=(\sqrt{A}/2)[\cosh{3\phi}+1/\cosh{3\phi}]$. If derived from
a perfect fluid or from the Born-Infeld Lagrangian
$\mathcal{L}_{\phi}$$ =-V_0 \sqrt{1+\varkappa^2 X}$, with
non-canonical $\rho=V_0/\sqrt{1-\varkappa^2
X},~P=-V_0\sqrt{1-\varkappa^2 X}, ~w=-1+\varkappa^2 X$, the
perturbations are adiabatic, and the effective sound speed
$c_s^2=-w(z)=c_a^2$. But, if derived from the canonical scalar field with
potential $V(\phi)$, entropic perturbations make the effective
sounds speed $c_s^2=1$. Agreeing in the static
limit, these three theories give the same adiabatic sound speed
$c_a^2=-w(z)$. Differing dynamically, they show different effective
sound speeds, $c_s^2=-w(z),~-w(z),~1$.

This toy model illustrates how the adiabatic sound speed depends only
on the equation of state, but the effective sound speed and growth of
fluctuations depend on dynamics. All three Chaplygin models can fit
the kinematic observations of the expansion history, but the
Born-Infeld model fails dynamically by providing insufficient power in
the observed large scale mass spectrum \cite{Amendola2003}.  This
failure can be remedied by generalizing the equation of state to
$P=-A/\rho^{\alpha}$ with $\alpha \sim 0$, so that this generalized
Chaplygin gas is nearly indistinguishable from a cosmological constant
\cite{Amendola2003,Bento,Zhu}.

\section{DARK GRAVITY: DYNAMICAL MODIFICATIONS OF GENERAL RELATIVITY} 

Because Dark Energy is contrived, requires fine tuning
and apparently cannot be tested in the laboratory or solar system, we
now turn to Dark Gravity as the alternative source of cosmological
acceleration. This Dark Gravity alternative arises naturally in braneworld
theories, naturally incorporates a classical extremely low spacetime
intrinsic curvature, and may unify "dark energy" and dark matter, and
possibly early and late inflation.  We will see how Dark Gravity, besides
being tested cosmologically, can also be tested in the solar system,
Galaxy or galaxy clusters.

\subsection{Present Local and Cosmological Tests of General Relativity} 

General Relativity is a rigid metric structure incorporating general
covariance (co-ordinate reparametrization invariance), the
Equivalence Principle, and the local validity of Newtonian gravity
with constant $G_N$, in the weak field and non-relativistic limits.
General covariance implies four local matter conservations laws
(Bianchi identities). The Einstein-Hilbert action is linear in Ricci
curvature, so that the Einstein field equations are second order,
the two tensorial (graviton) degrees of freedom are dynamic, but the
scalar and vector $g_{\mu\nu}$ degrees of freedom are constrained to
be non-propagating. Quantum gravity has always motivated high-curvature (Planck scale) modifications of
General Relativity.
Now, the surprising discovery of the accelerating universe motivates
extreme low-curvature (cosmological scale) modifications of General
Relativity.

General Relativity differs from Newtonian cosmology only by pressure
or relativistic velocity effects, which are tested in the solar
system, in gravitational lensing of light, in the primordial
abundance of light elements, in the dynamical age, and in the large
angular scale CMB and late-time mass power spectrum.  Therefore, in
order of linear scale, modifications of General Relativity must be
sought in:
\begin{itemize}
\item laboratory violations of
the Equivalence Principle (E$\ddot{o}$tvos experiments) and solar
system tests \cite{Damour} (lunar ranging, deflection of light,
anomalous orbital precessions of the planets, Moon
\cite{LueStarkman,Lue,Gabadadze}, secular increase in the
Astronomical Unit \cite{Iorio})
\item galaxy and galaxy cluster number counts \cite{IorioSecular}
\item gravitational weak lensing 
\item cosmological variation of Newton's $G_N$ and other "constants"
\item the suppression of fluctuation growth on large scales or
at late times.
\end{itemize}

\subsection{Classification of Modified Gravity
Theories by Spacetime Curvature of Their Vacuua} 

As already noted, in General Relativity only the tensor degrees of
freedom in the metric are propagating and the homogeneous RW
evolution depends only on $H(z)$.  If the Einstein-Hilbert action is
modified, additional scalar and vector degrees of freedom will
appear, and the evolution will also depend on $\dot{H},\ddot{H}$,
which can be represented by scalar or vector gravitational fields.
The basic distinction between high- and low-curvature modifications
of General Relativity depends on the spacetime (Ricci) curvature of
their vacua. It is simplest to begin by considering four-dimensional
metrical deformations of General Relativity, which are often inspired
by string-theory or M-theory 
\cite{DamourPolyakov,DamourPolyakov2} or
appear as
projections of higher-dimensional theories.

\subsubsection{Extra Degrees of Freedom in Four-Dimensional Gravity}

\bitem
\item Scalar-tensor gravity, the simplest and best-motivated extension
  of General Relativity \cite{Fujii,Cap2005}: In the original Jordon
  frame, a scalar gravitational field proportional to time-varying
  $1/G_N$, is linearly coupled to the Ricci scalar $R$. After a
  conformal transformation to the Einstein frame, the scalar
  gravitational field is minimally coupled to gravity, non-minimally
  coupled to matter.  In the Einstein frame, the gravitational field
  equations look like Einstein's, but the matter field is coupled to
  the scalar gravitational field as strongly as to the tensor
  gravitational field, so that test particles do not move along
  geodesics of the Einstein metric. Test particles move along
  geodesics of the original Jordon metric, so that the Weak
  Equivalence Principle holds.
  
  Scalar-tensor theories modify Einstein gravity at all scales and
  must be fine-tuned, in order to satisfy observational constraints.
  Nucleosynthesis and solar system constraints severely restrict any
  scalar field component, rendering any Dark Gravity effects on the
  CMB or $H(z)$ evolution imperceptible
  \cite{Bertotti,Cap2006,Catena}.
\item higher-order metric $f(R)$ theories: Stability of the
  equations of motion allows the Lagrangian to depend only on
  $R$, and only trivially on other curvature invariants,
  $P\equiv R_{\mu\nu} R^{\mu\nu},~Q\equiv R_{\alpha \beta \gamma
    \delta} R^{\alpha \beta \gamma \delta}$ \cite{Ostrogradski} or
  derivatives of any curvature scalar \cite{Woodard2006}.
  These $f(R)$ theories are equivalent to scalar tensor
  theories with vanishing Brans-Dicke parameter $\omega_{BD}=0$
  \cite{Teyssandier,Olmo,CapNojOdin}.

The simplest low-curvature modification
  \cite{Carroll,Carroll2,Nojiri}, replacing the Einstein Lagrangian
  density by $R -\mu^4/R$, leads to accelerated
  expansion at low-curvature $R \leq \mu^2\sim H_0^2$, but has
  negative kinetic energies and is tachyonically unstable.  This
  instability would be tolerable in empty space, but would be vastly
  and unacceptably amplified inside matter \cite{Dolgov}, and
  phenomenologically unacceptable outside matter \cite{Soussa}.
  These $f(R)$ theories, like their
  scalar-tensor gravity equivalents, can be fine-tuned to avoid these potential
  instabilities and satisfy supernova and solar system constraints
  \cite{Odintsov2,Odintsov3,Soussa,Woodard2006}, but not
  cosmological constraints \cite{Amendola2}.
\item TeVeS (relativistic MOND theory): Adding an additional vector
  gravitational field, could explain galactic rotation curves and the
  Tully-Fisher relation, without invoking dark matter, and possibly
  unify dark matter and ``dark energy'' \cite{Bek}. Because gravitons and
  matter have different metric couplings, TeVeS predicts that
  gravitons should travel on geodesics different from photon and
  neutrino geodesics, with hugely different arrival times from
  supernova pulses. It also predicts insufficient power in the third
  CBR acoustic peak \cite{Skordis}. In any case, now that WMAP data
  requires dark matter \cite{Spergel}, the motivation for TeVeS
  disappears.
\eitem

\subsubsection{Extra Dimensional Modifications of Einstein Gravity} 

In extra dimensional braneworld theories, scalar fields appear
naturally as dilatons and modify Einstein gravity at high-
curvature, by brane warping \cite{Randall,Binutray}, or at low-
curvature, by brane leakage of gravity \cite{DGP}.  If quantized,
these theories encounter serious theoretical problems (ghosts,
catastrophic ultra-violet instabilities, strong coupling problems).
Until these problems can be overcome, these theories can only be
regarded as effective field theories, incorporating an extremely low
{\em infra-red} scale at low spacetime curvature. This suggests an
infra-red/ultra-violet connection, since effective field theories
ordinarily incorporate {\em ultra-violet} parameters.

In the original DGP model \cite{DGP,Deffayet,Deffayet2}, the brane's
finite stiffness leads to an effective modified Friedmann equation,
\bea
H^2+k/a^2-H/r_c=H^2+k/a^2-H H_0/\beta=\varkappa^2\rho/3\\
(H(z)/H_0)^2=[1/2\beta+\sqrt{(1/2\beta)^2+\Omega_{m0}
  (1+z)^3}]^2+\Omega_{K0}(1+z)^2,\quad 1 \equiv \Omega_{m0}+
\Omega_{K0} +\sqrt{1-\Omega_{K0}}/\beta \eea on the four-dimensional
brane by inducing an additional curvature term $H/r_c$ at the
cosmological scale $\beta\equiv H_0 r_c=1.39,~r_c= \beta/H_0 \equiv
H_{\infty}^{-1} \sim 5.7
~Gpc$. This modified Friedmann equation interpolates between
Einstein's pressure-free universe at large redshifts, and the empty
de Sitter universe with constant Hubble expansion $H_{\infty}\equiv 
1/r_c=H_0/\beta$, in the
asymptotic future.  The universe began its late acceleration at
$z_{acc}=(2\Omega_{m0}/\beta^2)^{1/3}-1 \sim 0.58$.  This is the
original DGP model fit on the fourth line of Table I, which turns out to
be spatially practically flat.

In flat 3-space, the modified Friedmann equation has the
self-accelerating solution \bea
H=\sqrt{\varkappa^2\rho/3+(1/2r_c)^2}+1/2r_c\\
H(z)/H_0=1/2\beta+\sqrt{(1/2\beta)^2+\Omega_{m0} (1+z)^3},\qquad 1
\equiv \Omega_{m0}+1/\beta, \eea Because this self-accelerating
solution has a Ricci-curved vacuum, Birkhoff's Theorem does not apply
on the 4D brane, and Einstein gravity still holds at the shortest
distances. However, about any isolated condensation of Schwarzchild
radius $r_S\equiv2 G_N M/c^2$, the self-accelerating metric \be
g_{tt}=1-r_S/2r+\sqrt{r_S^2 r/2 r_{\star}^3},\qquad
g_{rr}^{-1}=1+r_S/2 r-\sqrt{r_S^2 r/8 r_{\star}^3},\qquad r\lesssim
r_{\star}, \ee and Einstein gravity already breaks down at the
Vainshtein scale \cite{DGZ} defined by \be r_{\star}\equiv (r_S r_c^2)^{1/3}\sim
(H_0 r_S)^{1/3} H_0^{-1}\ll H_0^{-1}. \ee This
scale,  surprisingly intermediate between $r_S$ and $H_0^{-1}$, is also where the growth of fluctuations
changes from Einstein gravity to linearized DGP or scalar-tensor
Brans-Dicke growth, with an
effective Newton's constant slowly decreasing by no more than a factor
two \cite{LueScoccimaro}.

The original flat DGP model can be generalized \cite{DvaliTurn} to \be
H^2-H^{\alpha} r_c^{\alpha-2}=\varkappa^2\rho/3,\qquad 1/\beta \equiv
(1-\Omega_{m0})^{1/(2-\alpha)},\qquad
1=\Omega_{m0}+\beta^{\alpha-2},\ee which is equivalent to a "dark
energy" $\rho_{DE}\equiv 3 M_P^2 H^2-\rho=3 M_P^2 H^{\alpha}
r_c^{\alpha-2},~w_{DE}=-1+\alpha/2$. This generalization reduces to
the original flat DGP form for $\alpha=1$, but otherwise interpolates
between the $\Lambda$CDM model for $\alpha=0,~\beta=1.18$ and the
Einstein-de Sitter model for $\alpha=2,~\beta=\infty $. For small
$\alpha$, it describes a slowly varying cosmological constant.
Excluding the CMB shift data, the joint SNLS-BAO data fits this
nearly static Dark Gravity model fit with $\alpha=-0.17^{+0.87}_{-0.63},~\beta=1.16$
or $r_c=\beta/H_0 \sim 8~Gpc$ \cite{Fairbairn}.  This is
the generalized flat DGP model on the last line of Table I.

\begin{figure}[top]
  \begin{center} 
  \includegraphics[bb=0 0 628 546,width=1.23 \textwidth]{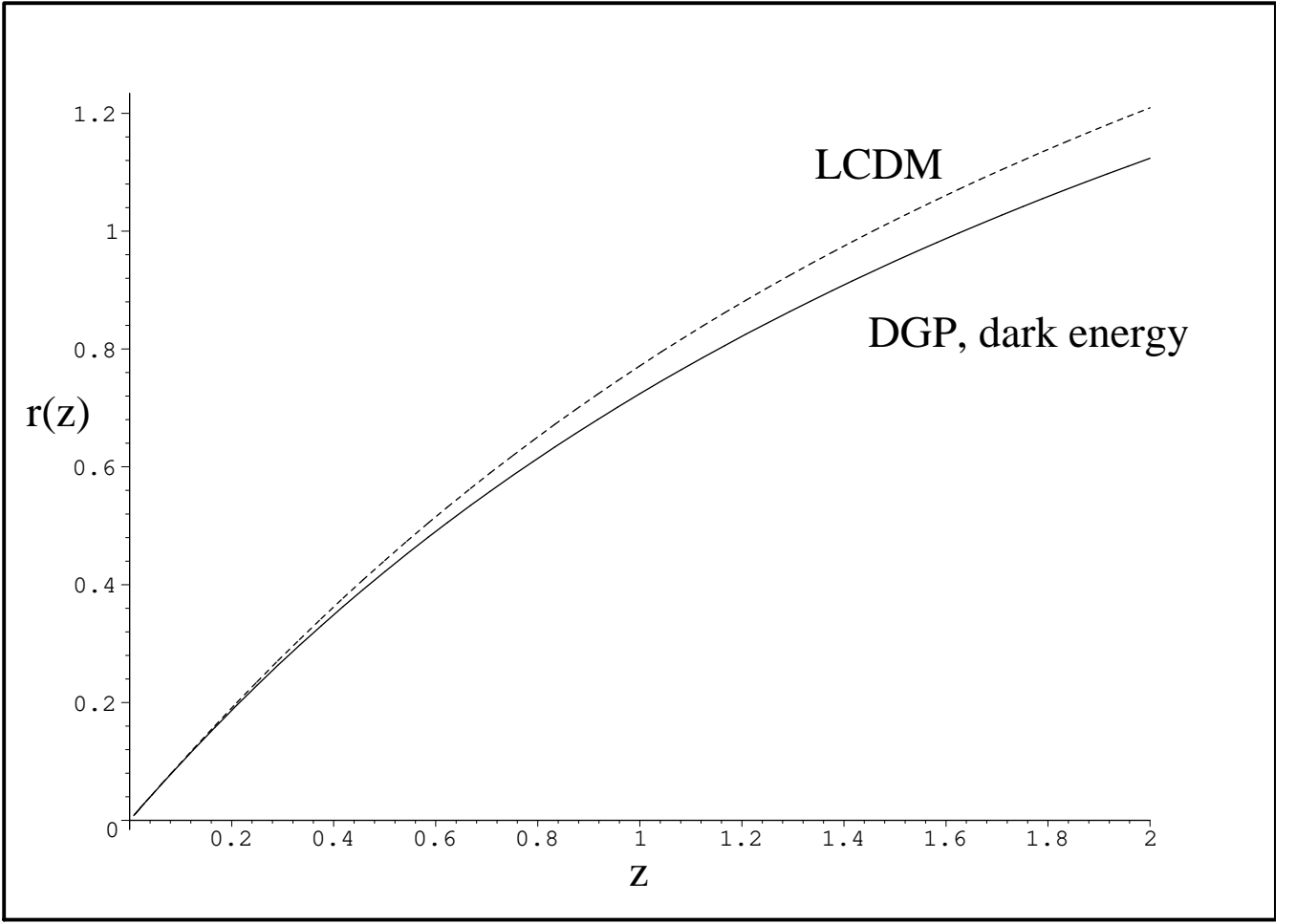}
  \caption{The proper motion or conformal distance $r(z)=c\int_0^z
    dz'/H(z')$ back to redshift $z$, calculated for flat $\Lambda$CDM
    (dashed) and DGP Dark Gravity (solid) models with present matter
    fraction $\Omega_{m0}=0.3$.  The DGP Dark Gravity model is also
    mimicked by a Dark Energy model with $w(a)=-0.78+0.32 z/(1+z)$.
    Between static and dynamic ``dark energy'', the differences in
    distances and in expansion history $H(z)=c dz/r(z)$ are small, but
    the differences in growth factor will be larger in FIG. 4. (from
    \cite{Koyama}).}
  \end{center}
\end{figure}

\begin{figure}[top]
  \begin{center} 
  \includegraphics[bb=0 0 628 546,width=1.23 \textwidth]{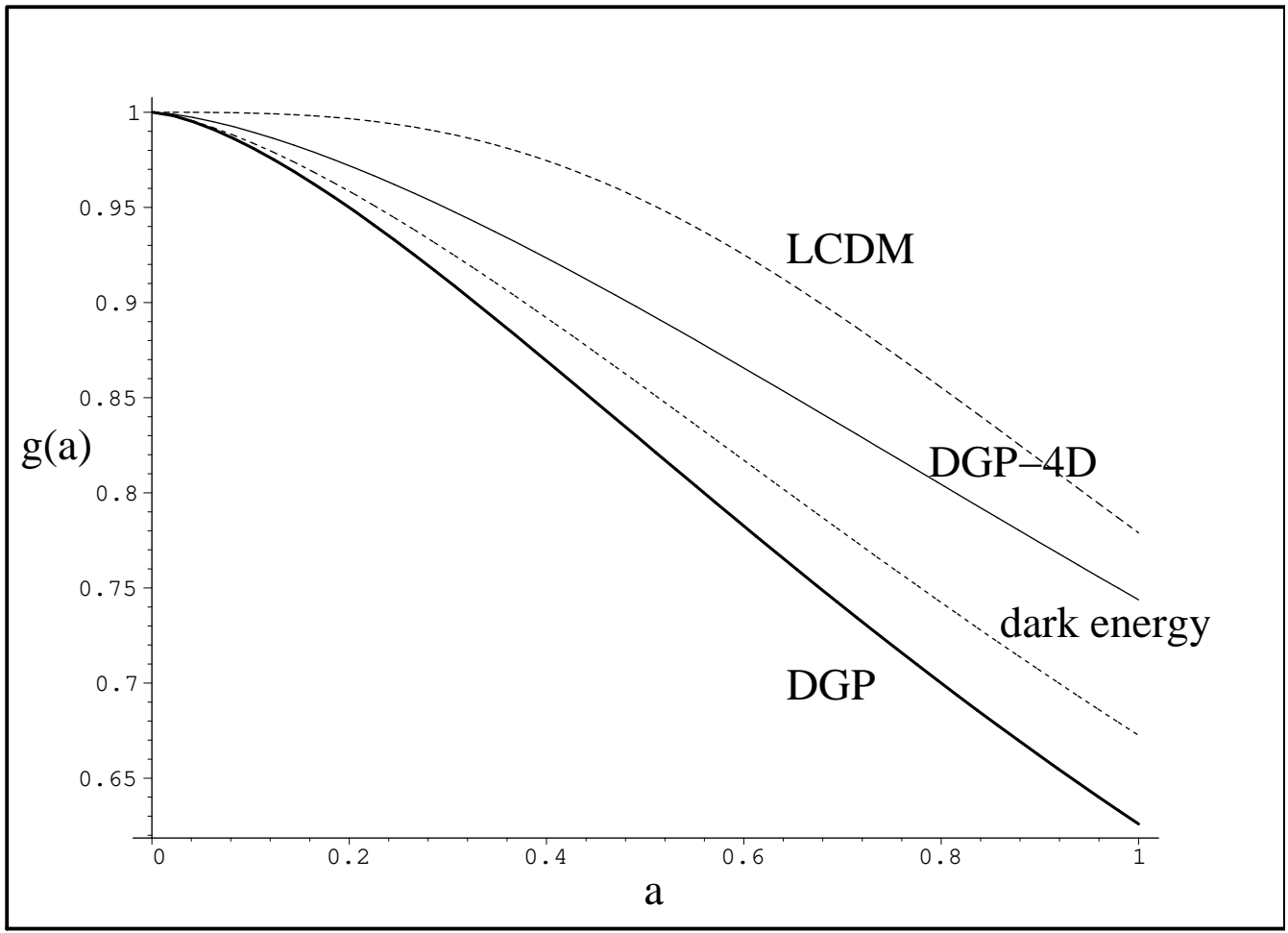}
  \caption{The linear growth history $g(a)\equiv \delta/a$ for flat $\Lambda$CDM
    (long dashed), DGP Dark Gravity (thick solid) and Dark Energy
    models in Fig. 3. Because the DGP model Newton's ``constant''
    weakens with time, it shows a little more growth suppression than that in
    the mimicking Dark Energy model. DGP-4D (thin solid) shows the
    incorrect result obtained by neglecting perturbations of the DGP
    5D Weyl tensor. The 5D Weyl tensor perturbations
    make dynamical Dark Gravity (DGP) and Dark Energy hard to resolve,
    but both dynamical models are distinguishable 
    from static Dark Energy. (from \cite{Koyama})}.
\end{center}
\end{figure}

\subsection{Prospective Tests of Modified Gravity}
{\em Prospective cosmological observations:} Figure 3 compares the
recent expansion histories Koyama and Maartens \cite{Koyama}
calculated in the static flat $\Lambda$CDM model with the dynamic flat
DGP Dark Gravity model $\beta=1.41$, chosen to fit
$\Omega_{m0}=0.3$, and with its Dark Energy mimic.  The degeneracy in
these three models is lifted by their different linear growth
factors plotted in Figure 4: DGP Dark Gravity always suppresses
growth a little more than Dark Energy does, but substantially more
than the Concordance Model.  The present normalized linear growth factors
$g(a)=\delta/a=0.61, 0.68.0.80$ for DGP Dark Gravity, Dark Energy,
$\Lambda$CDM respectively will lead to slightly different large-scale CMB and
gravitational weak lensing (cosmic shear) convergence effects.

These linear growth factors have been calculated \cite{Koyama} on
subhorizon scales and agree with \cite{LueStarkman,Ishak}, but are not yet
reliable on superhorizon scales, where gravity leakage is most
important. Unfortunately,
the large-scale CMB angular power spectrum
is obscured by cosmic variance and by foreground effects of
nearby structures. Nevertheless, the linear growth factors already suggest
that next-generation observations may distinguish static from
dynamic "dark energy", but will be unable to distinguish Dark Energy
from DGP Dark Gravity. To do so will require a newer, more ambitious
weak lensing shear survey \cite{Ishak}.

{\em Prospective solar system and galaxy cluster observations:} 
The modifications to Einstein gravity at the Vainstein intermediate
scale $r_{*}$ may also be tested in next-generation solar system
measurements of anomalous precessions of planetary or lunar orbits
\cite{LueStarkman,DGZ} or of a secular increase in the Astronomical
Unit \cite{IorioSecular}. These Vainstein scale modifications may
also be observable in precision tests about other isolated stars
($r_S \sim 3~km, ~ r_{\star} \sim 280~pc$) or about isolated spherical galaxy
clusters $r_S \sim 100~pc, ~ r_{\star} \sim 28~Mpc$) \cite{Iorio}.

\section{CONCLUSIONS: $\Lambda$CDM, FINE-TUNED DARK ENERGY, OR MODIFIED
GRAVITY} 

We have reviewed present and prospective observations of ``dark
energy'' as the source of the observed late cosmological
acceleration, in order to emphasize the differences between kinematical and
dynamical observations, between static and dynamic ``dark energy'',
and between Dark Energy and Dark Gravity.
We conclude that

\begin{itemize}
\item Cosmological acceleration is explicable by either a small
  fine-tuned cosmological constant or by ``dark energy'', which is now
  at most moderately dynamic.  This ``dark energy'', if dynamic, is
  either an additional, ultra-light matter constituent within General
  Relativity, or a low-curvature modification of Einstein's field
  equations.
\item The best and simplest fit to the expansion history, static "dark
  energy" ($\Lambda$CDM), interprets the cosmological constant as a
  classical intrinsic geometric Ricci curvature, rather than as vacuum
  energy.  This side-steps the cosmological constant problems, why
  quantum vacuum fluctuations apparently do not gravitate and why the
  current matter density is roughly that of ``dark energy''.
\item The homogeneous expansion history can also be fitted by
  moderately dynamical "dark energy". Only observing the large-scale
  inhomogeneity growth rate will distinguish between dynamic and
  static "dark energy".
\item Projected cosmological observations of the growth factor in the
  large-scale angular power spectrum, mass power
  spectrum, or 
gravitational weak lensing convergence should distinguish
  static from dynamic "dark energy", but not Dark Energy from
  Dark Gravity.
\end{itemize}

Originally invoked to explain late cosmological acceleration within
General Relativity, quintessence and k-essence Dark Energy
fail to explain the
Cosmological Coincidence ``Why Dark Energy appears now?'', without fine-tuning or anthropic reasoning.  Low-
curvature modifications of Einstein gravity, such as DGP, are
conceptually less contrived than finely-tuned Dark Energy, and arise naturally
in braneworld theories. They explain
cosmological acceleration as a natural consequence of geometry, may
unify early and late inflation,, and may even be tested by refined 
solar system or galaxy observations.

\begin{acknowledgments}
I thank Richard Woodard (University of Florida), for helpful
discussions of Ostrogradski's Theorem and $f(\mathcal{R})$ theories,
and Dallas Kennedy (MathWorks) and Roy Maartens (Portsmouth), for
critical comments.
\end{acknowledgments}
\bibliography{bibliographyPhysrev}
\end{document}